\documentclass[12pt]{article}
\usepackage{fancyhdr}
\usepackage{amsbsy}
\usepackage{amssymb}

\usepackage{epsfig}

\newcommand\comma{\ ,\ \ }

\begin{document}

\title{Modular Cosmology, Thermal Inflation, Baryogenesis\\
and Predictions for Particle Accelerators}

\author{
Donghui Jeong$^1$, Kenji Kadota$^2$, Wan-Il Park$^1$ and Ewan D. Stewart$^1$ \\[2ex]
$^1$ {\em Department of Physics, KAIST, Daejeon 305-701, South Korea}\\
$^2$ {\em Department of Physics, University of California, Berkeley, CA 94720, USA}
}

\maketitle

\begin{abstract}
Modular cosmology is plagued by overproduction of unwanted relics, gravitinos and especially moduli, at relatively low energy scales.
Thermal inflation provides a compelling solution to this moduli problem, but invalidates most baryogenesis scenarios.
We propose a simple model in which the MSSM plus neutrino mass term $(LH_u)^2$ is supplemented by a minimal flaton sector to drive the thermal inflation, and make two crucial assumptions: the flaton vacuum expectation value generates the $\mu$-term of the MSSM and $m_L^2 + m_{H_u}^2 < 0$.
The second assumption is particularly interesting in that it violates a well known constraint, implying that there exists a nearby deep non-MSSM vacuum, and provides a
clear signature of our model which can be tested at future particle accelerators. We show that our model leads to thermal inflation followed by Affleck-Dine leptogenensis along the $LH_u$ flat direction.
A key feature of our leptogenesis scenario is that the $H_uH_d$ flat direction is also induced to temporarily acquire a large value, playing a crucial role in the leptogenesis, as well as dynamically shielding the field configuration from the deep non-MSSM minimum, ensuring that the fields relax into our MSSM vacuum.

\end{abstract}

\thispagestyle{fancy}
\rhead{KAIST-TH/2004-07\\UCB-PTH-04/18}

\newpage

\section{Introduction} \label{intro}

Among the many unresolved problems in cosmology, the origin of the matter-antimatter asymmetry, or baryon asymmetry, is the one we can most easily recognize because we could not exist without it.
According to recent data from the WMAP experiment \cite{Spergel}, the asymmetry is
\begin{equation}
\frac{n_b}{n_{\gamma}} = (6.5 \pm 0.4) \times 10^{-10}.
\end{equation}
Since the time, about forty years ago, when Sakharov showed that three conditions must be satisfied to generate this asymmetry \cite{Sakharov}, many authors have tried to explain this asymmetry with many different models.
But we still don't have any idea which, if any, is the right one because baryogenesis usually depends on unknown physics beyond the reach of current experiments.

Besides producing necessary relics such as the baryon asymmetry, a successful cosmological history must also avoid or dilute unwanted cosmological relics such as gravitinos \cite{KhLi} and moduli \cite{CFKRR} which can destroy the successful predictions of Big-Bang nucleosynthesis \cite{KT} or over-close the universe, depending on the scale of supersymmetry breaking.
In gravity mediated supersymmetry breaking scenarios, these particles have masses of order $m_\mathrm{EW} \sim 10^2$ to $10^3$ GeV and Plank scale suppressed couplings.
This indicates their long life time and possibility of disturbing Big Bang nucleosynthesis.
Both particles may be thermally produced in dangerous abundances in the early universe, but such production might be avoided if the reheat temperature after inflation was sufficiently low.
More seriously, moduli have another more dangerous source, production of coherent moduli oscillations with amplitude of order of the Plank scale when the Hubble parameter drops below the vacuum mass of the moduli \cite{CFKRR}.
In the case of gravity mediated supersymmetry breaking scenarios, this occurs at an energy scale of the order of the intermediate scale, $\sqrt{m_\mathrm{EW} M_\mathrm{Pl}} \sim 10^{10}$ to $10^{11}$ GeV, i.e. after, or at best at, the end of the usual high energy inflation.
Thus, we face a disaster unless there is a proper amount of late time entropy release.

Fortunately, thermal inflation \cite{Stewart,DFN} provides a compelling solution to this moduli problem.
It is expected to occur after the usual high energy inflation, at an energy scale between the intermediate scale and the electroweak scale, releasing enough entropy to dilute the unwanted relics.
It is driven by the potential energy of a `flaton' field which is held at the origin by thermal effects.
When the temperature drops sufficiently, the flaton rolls away from the origin, ending the thermal inflation.
The flaton then oscillates about its vacuum expectation value of the order of $10^{10}$ to $10^{12}$ GeV till its decay completes at a temperature of order GeV.

There are several categories of baryogenesis models and it is important to consider them in the context of the above discussion of a complete and consistent cosmological history.
One is baryogenesis by out of equilibrium decay of heavy particles.
For example, GUT baryogenesis \cite{Yo} uses particles with mass of the order of the GUT scale, $M_\mathrm{GUT} \sim 10^{16}$ GeV.
Right-handed neutrino leptogenesis \cite{FuYa} uses decay of heavy right-handed neutrinos of mass $10^{12}$ to $10^{15}$ GeV, and anomalous $B+L$ violating electroweak processes mediated by sphalerons \cite{KRS} to convert the generated lepton asymmetry to baryon asymmetry.
However, in both cases, very high energy scales are required to produce the heavy particles.
This means the generated asymmetry would be diluted to negligible amounts by thermal inflation or any other form of entropy production required to dilute the unwanted relics that are produced at lower energy scales.

Another category is electroweak baryogenesis \cite{Trodden}.
It uses the electroweak phase transition and sphalerons.
The attractive point of this mechanism is that it works at a low energy scale, the electroweak scale, so it is rather independent of unknown high energy physics.
However, it seems that the window in the Minimal Supersymmetric Standard Model (MSSM) for this mechanism is small \cite{CEKKLW}.
Moreover, assuming thermal inflation, the reheat temperature after flaton decay is too low even for this mechanism.

The last major category, Affleck-Dine (AD) baryogenesis \cite{Affleck_Dine} is distinctive from the previous two categories due to its efficiency.
Using coherent fields along MSSM flat directions, it can generate huge asymmetries which can survive substantial dilution.
Indeed, in the case of gauge mediated supersymmetry breaking, it can even produce a large enough asymmetry to survive the full entropy production of the thermal inflation needed to dilute the moduli to acceptable levels \cite{GMM}.
However, especially in the case of gauge mediated supersymmetry breaking, the AD condensate can fragment into Q-balls \cite{Coleman,KKTak}, though it is not clear how serious a problem this is as the absorbed baryon asymmetry may be released through the decay of the Q-ball \cite{Qdecay}.
In the case of gravity mediated supersymmetry breaking scenarios, the required entropy production is greater, so that not even AD baryogenesis can survive the full entropy production of the thermal inflation.

Thus, if we assume supersymmetry breaking is mediated by gravitational strength interactions and thermal inflation solves the moduli problem, all the baryogenesis models suggested so far do not work.
Therefore, a new scenario of baryogenesis is required under these assumptions.
This is our motivation.
The new baryogenesis model must satisfy two requirements.
The first is that, as it cannot work either before the thermal inflation or after the flaton decay, it must work around the end of the thermal inflation.
The other is that it must generate a large asymmetry which can survive the dilution by the entropy released in the flaton decay.
In this paper, we suggest such a model.
It is similar to an earlier model by one of us \cite{SKY}, but is based on a more minimal model and crucially relies on some dynamics neglected in \cite{SKY}.

We begin our discussion by explaining the construction of our model in Section~\ref{model}.
In Section~\ref{Baryo}, we describe our baryogenesis scenario.
In Section~\ref{CnS}, we discuss the constraints on parameters and the stability of dangerous field directions.
In Section~\ref{Sim}, we show the results of numerical simulations of the field dynamics.
In Section~\ref{after}, we discuss some of the issues involved in determining whether our generated asymmetry is preserved.
In Section~\ref{dis}, we conclude and comment about future work.

\section{The Model} \label{model}

\subsection{The superpotential} \label{Pot}

The MSSM superpotential is
\begin{equation}\label{mssm}
W = \lambda_u^{ij} \epsilon_{ab} Q_i^{a\alpha} H_u^b u_j^\alpha
+ \lambda_d^{ij} \epsilon_{ab} Q_i^{a\alpha} H_d^b d_j^\alpha
+ \lambda_e^{ij} \epsilon_{ab} L_i^a H_d^b e_j
+ \mu \epsilon_{ab} H_u^a H_d^b
\end{equation}
where $i$ is the generation index, $a$ is the SU(2) index and $\alpha$ is the SU(3) index.

Neutrino oscillations \cite{BP,SNO} imply nonzero neutrino masses.
So we include the following Majorana neutrino mass term
\begin{equation}\label{numass}
\frac{1}{2} \lambda_\nu^{ij} \epsilon_{ab} L_i^a H_u^b \epsilon_{cd} L_j^c H_u^d.
\end{equation}

We expect thermal inflation as a solution of the moduli problem.
So we need a flaton sector.
For simplicity, we consider only a minimal flaton sector\footnote{One interesting alternative would be to embed a global U(1) Peccei-Quinn symmetry \cite{PQ} in the flaton sector.
Our scenario would work similarly in that case, though there is a danger of overproduction of axions \cite{JEKim} by the flaton decay.}
consisting of a single field $\phi$ with self-interaction
\begin{equation}
\frac{1}{4} \lambda_\phi \phi^4.
\end{equation}
This naturally stabilizes the flaton field at a vacuum expectation value of the order of $10^{10}$ to $10^{12}$ GeV.
The flaton must also have some interactions with other fields, for example a coupling of the form $\lambda_\chi \phi \bar{\chi} \chi$, to provide the finite temperature potential which holds it at the origin during the thermal inflation.
$\chi$ and $\bar{\chi}$ may carry MSSM charges but, as they become very heavy at the end of thermal inflation, are not MSSM fields.
They should form complete SU(5) representations to preserve gauge coupling unification, but would still be expected to leave an imprint on the renormalization of MSSM parameters from the GUT scale.
As these fields and couplings do not affect our dynamics beyond holding the flaton at the origin during the thermal inflation, we do not specify them explicitly.

The origin of $\mu$ in Eq.~(\ref{mssm}) is unknown.
We assume it is absent initially and generated dynamically by the vacuum expectation value of the flaton field via the interaction
\begin{equation}
\lambda_{\mu} \phi^2 \epsilon_{ab} H_u^a H_d^b
\end{equation}
as was suggested long ago in Ref.~\cite{JH}.
In our vacuum, this coupling gives
\begin{equation}
\mu = \langle0| \lambda_\mu \phi^2 |0\rangle
\end{equation}
which has the correct size for $\mu$ under the assumption
\begin{equation}
|\lambda_\mu| \sim |\lambda_\phi|
\end{equation}
and flaton mass of electroweak scale.
This coupling also gives a reheat temperature after flaton decay high enough for thermalization and standard freeze out of neutralino dark matter \cite{SKY}.

Given the structure of our model, we could also expect a term
\begin{equation}
\frac{1}{2} \lambda_H \left( \epsilon_{ab} H_u^a H_d^b \right)^2
\end{equation}
with
\begin{equation}
\left| \lambda_H \right| \sim \left| \lambda_\mu \right|
\sim \left| \lambda_\phi \right|.
\end{equation}
However, we assume
\begin{equation}
|\lambda_\nu| \gg |\lambda_\phi|
\end{equation}
to give correct values for the neutrino masses and an appropriate scale for thermal inflation.
As we shall see, this means that
$|\lambda_H \epsilon_{ab} H_u^a H_d^b| \ll m_\mathrm{EW}$
throughout our dynamics, so such a term is negligible.

Thus, our superpotential is
\begin{eqnarray}\label{W}
W & = & \lambda_u^{ij} \epsilon_{ab} Q_i^{a\alpha} H_u^b u_j^\alpha + \lambda_d^{ij} \epsilon_{ab} Q_i^{a\alpha} H_d^b d_j^\alpha + \lambda_e^{ij} \epsilon_{ab} L_i^a H_d^b e_j \nonumber \\
&& \mbox{} + \lambda_\mu \phi^2 \epsilon_{ab} H_u^a H_d^b + \frac{1}{2} \lambda_\nu^{ij} \epsilon_{ab} L_i^a H_u^b \epsilon_{cd} L_j^c H_u^d + \frac{1}{4} \lambda_\phi \phi^4.
\end{eqnarray}

We could use a $\mathbb{Z}_4$ symmetry with charges
\begin{equation} \label{syma}
\mathbb{Z}_4 \{H_u,H_d,Q,u,d,L,e,\phi\} = \{0,2,2,2,0,0,2,1\}
\end{equation}
plus the usual $R$-parity, to enforce the above form for the superpotential.
This $\mathbb{Z}_4$ symmetry could be a gauge symmetry if it is anomaly free \cite{Ross}.
It can be made anomaly free by adding an extra two SU(5) $\mathbf{10}$ representations and two SU(5) $\mathbf{\bar{10}}$ representations, with coupling $\lambda_\chi^{ij} \phi \mathbf{\bar{10}}_i \mathbf{10}_j$ where $i = 1,2$ and $j = 1,\ldots,5$.
Some such coupling would in any case be necessary for the thermal inflation as already mentioned.
Gauging the $\mathbb{Z}_4$ symmetry is one way to avoid the potential domain wall problem \cite{JEKim} caused by the flaton potential's four minima.
We assume we are free from that problem.

\subsection{Our ansatz}

In our vacuum, all the MSSM supersymmetric flat directions in field space are stable at the origin.
However, during thermal inflation the flaton is at the origin, and so there is no $\mu$-term.
This may cause some supersymmetric flat directions involving $H_u$ or $H_d$ to be unstable during thermal inflation.
These potentially unstable flat directions are $H_uH_d$ and $L_iH_u$.
Thus, we assume that the only fields which get nonzero expectation values are $H_u$, $H_d$, $L_i$ and $\phi$.
We will check the consistency of this ansatz in Section~\ref{CnS}.

For simplicity, we will truncate to a single generation.
This may be essentially correct if all the matrices of leptonic sector are simultaneously diagonalizable, while otherwise will hopefully still capture the essential points.

Gauge fixing and imposing the $D$-term constraints for these flat directions, we can parameterize them as
\begin{equation}
H_u =
\left(\begin{array}{c} h_u \\ 0
\end{array}\right) \comma H_d = \left(\begin{array}{c} 0 \\ h_d
\end{array}\right) \comma L = \left(\begin{array}{c} 0 \\ l
\end{array}\right) \comma \phi = \phi
\end{equation}
with the remaining $D$-term constraint
\begin{equation}\label{D}
D = |h_u|^2 - |h_d|^2 - |l|^2 = 0
\end{equation}
and corresponding gauge degree of freedom. Integrating out the remaining gauge field and choosing the gauge $A_\mu = 0$ gives
\begin{equation}
j_\mu = \frac{1}{2i} \left( h_u^* \partial_\mu h_u - h_u \partial_\mu h_u^* \right)
- \frac{1}{2i} \left( h_d^* \partial_\mu h_d - h_d \partial_\mu h_d^* \right)
- \frac{1}{2i} \left( l^* \partial_\mu l - l \partial_\mu l^* \right) = 0.
\end{equation}
Thus, the superpotential reduces to
\begin{equation}\label{Wred}
W = \lambda_\mu \phi^2 h_u h_d + \frac{1}{2} \lambda_\nu l^2 h_u^2 + \frac{1}{4} \lambda_\phi \phi^4
\end{equation}
and the $F$ and $D$-term parts of the potential to
\begin{eqnarray}\label{Fterm}
V_F & = & \left| \lambda_\mu \phi^2 h_d + \lambda_\nu l^2 h_u \right|^2
+ \left| \lambda_\mu \phi^2 h_u \right|^2
+ \left| \lambda_\nu l h_u^2 \right|^2
+ \left| 2 \lambda_\mu \phi h_u h_d + \lambda_\phi \phi^3 \right|^2,
\end{eqnarray}
\begin{equation} \label{Dterm}
V_D = \frac{1}{2} g^2 \left( |h_u|^2 - |h_d|^2 - |l|^2 \right)^2
\end{equation}
where $g^2 = ( g_1^2 + g_2^2 ) / 4$. The vacuum supersymmetry
breaking potential is
\begin{eqnarray}\label{vsb}
V_\mathrm{vsb} & = & V_0 + m_{H_u}^2 |h_u|^2 + m_{H_d}^2 |h_d|^2 + m_L^2 |l|^2 + m_\phi^2 |\phi|^2 \nonumber \\
&& \mbox{} + \left[ A_\mu \lambda_\mu \phi^2 h_u h_d + \frac{1}{2} A_\nu \lambda_\nu l^2 h_u^2 + \frac{1}{4} A_\phi \lambda_\phi \phi^4 + \mbox{c.c.} \right]
\end{eqnarray}
where $V_0$ is adjusted to give zero cosmological constant, and all the other supersymmetry breaking parameters are of the order of the electroweak scale.

Again for simplicity, we neglect any field dependent renormalization of the supersymmetry breaking parameters.
Note that taking into account the $\phi$ dependent renormalization of the supersymmetry breaking parameters will be important for accurate quantitative comparison of our constraints on the supersymmetry breaking parameters near the end of thermal inflation with those parameters in our vacuum.

Thus, the whole potential for these fields is given by
\begin{equation}\label{Vwhole}
V=V_\mathrm{vsb} + V_F + V_D.
\end{equation}

\section{Baryogenesis} \label{Baryo}

We are now in a position to describe how baryogenesis can occur in
our model.

\subsection{Before the end of thermal inflation} \label{dyna}

At high temperature, the thermal masses hold all the fields at the origin. So, we take our initial condition as thermal inflation with all fields having zero expectation value.
We assume $m_\phi^2 < 0$ and $m_L^2 + m_{H_u}^2 < 0$ so that the flaton $\phi$ and the supersymmetric flat direction $LH_u$ become unstable as the temperature drops during thermal inflation.
We assume $LH_u$ rolls away first.
Once a field rolls away, it decouples from the thermal bath, and we neglect any remaining thermal potential.
We will discuss the wider implications of our rather non-trivial assumption that $m_L^2 + m_{H_u}^2 < 0$ in Section~\ref{CnS}.

While the flaton is still held at the origin, the potential for $l$ and $h_u$ is
\begin{eqnarray} \label{lhupot}
V & = & V_0 + m_{H_u}^2 |h_u|^2 + m_L^2 |l|^2
+ \left[ \frac{1}{2} A_\nu \lambda_\nu l^2 h_u^2 + \mbox{c.c.} \right] \nonumber \\
&& \mbox{}
+ \left| \lambda_\nu l^2 h_u \right|^2
+ \left| \lambda_\nu l h_u^2 \right|^2
+ \frac{1}{2} g^2 \left( |h_u|^2 - |l|^2 \right)^2
\end{eqnarray}
which has minimum at
\begin{equation}\label{lhuph}
A_\nu \lambda_\nu l^2 h_u^2 = - \left| A_\nu \lambda_\nu l^2 h_u^2 \right|
\end{equation}
and
\begin{equation}\label{huti}
|h_u|^2 \simeq \frac{|A_\nu| + \sqrt{|A_\nu|^2 - 6 \left( m_L^2 + m_{H_u}^2 \right)}\,}{6 |\lambda_\nu|} + \frac{\left( m_L^2 - m_{H_u}^2 \right)}{4 g^2},
\end{equation}
\begin{equation}\label{lti}
|l|^2 \simeq \frac{|A_\nu| + \sqrt{|A_\nu|^2 - 6 \left( m_L^2 + m_{H_u}^2 \right)}\,}{6 |\lambda_\nu|} - \frac{\left( m_L^2 - m_{H_u}^2 \right)}{4 g^2}.
\end{equation}

For simplicity, we assume $l$ and $h_u$ settle down to this minimum before $\phi$ or $h_d$ start rolling away from the origin.
Note that substituting Eqs.~(\ref{huti}) and~(\ref{lti}) into Eq.~(\ref{Dterm}) gives a contribution $-(m_L^2 - m_{H_u}^2)/2$ to $h_d$'s mass squared.
We assume that this does not destabilize $h_d$.
A sufficient condition for this is
\begin{equation}
m_{H_u}^2 + m_{H_d}^2 > \frac{1}{2} \left( m_L^2 + m_{H_u}^2 \right)
\end{equation}
though it is probably not necessary.

\subsection{At the end of thermal inflation}
\label{end}

Thermal inflation ends when $\phi$ rolls away from the origin.
We assumed $|\lambda_\nu| \gg |\lambda_\phi|$ to give correct values for the neutrino masses and an appropriate scale for thermal inflation.
This condition also ensures that the flaton dynamics is dominant as the flaton gets a larger value than $h_u$, $h_d$ or $l$.
Thus, the flaton can be regarded in isolation, neglecting any back reaction from the other fields.
See Eq.~(\ref{energy}).
The pure flaton part of the potential is
\begin{eqnarray}\label{VTI}
V_\mathrm{TI} & = & V_0 + m_\phi^2 |\phi|^2
+ \left[ \frac{1}{4} A_\phi \lambda_\phi \phi^4 + \mbox{c.c.} \right]
+ \left| \lambda_\phi \phi^3 \right|^2
\end{eqnarray}
which has minima at $\phi = \phi_0$ plus $\mathbb{Z}_4$ symmetric points, where
\begin{equation}
A_\phi \lambda_\phi \phi_0^4 = - \left| A_\phi \lambda_\phi \phi_0^4 \right|
\end{equation}
and
\begin{equation}
|\phi_0|^2 = \frac{\sqrt{- 12 m_\phi^2 + |A_\phi|^2}\, + |A_\phi|}{6 |\lambda_\phi|}.
\end{equation}
From Eqs.~(\ref{Wred}), (\ref{Fterm}) and~(\ref{vsb}), once the flaton settles down, the MSSM parameters $\mu$ and $B$ will be given by
\begin{equation}\label{mu}
\mu = \lambda_\mu \phi_0^2
\end{equation}
and
\begin{equation}\label{B}
B = A_\mu + \frac{2 \lambda_\phi^* {\phi_0^*}^4}{|\phi_0|^2} = A_\mu + \frac{2 \lambda_\phi^* \lambda_\mu^2 {\mu^*}^2}{|\lambda_\mu^3 \mu|}.
\end{equation}

The remainder of the potential,
\begin{eqnarray}\label{VAD}
V_\mathrm{AD} & = & m_{H_u}^2 |h_u|^2 + m_{H_d}^2 |h_d|^2 + m_L^2 |l|^2 \nonumber \\ && \mbox{}
+ \left[ A_\mu \lambda_\mu \phi^2 h_u h_d + \frac{1}{2} A_\nu \lambda_\nu l^2 h_u^2 + 2 \lambda_\phi^* \phi^{*3} \lambda_\mu \phi h_u h_d + \mbox{c.c.} \right] \nonumber \\ && \mbox{}
+ \left| \lambda_\mu \phi^2 h_d + \lambda_\nu l^2 h_u \right|^2
+ \left| \lambda_\mu \phi^2 h_u \right|^2
+ \left| \lambda_\nu l h_u^2 \right|^2
+ \left| 2 \lambda_\mu \phi h_u h_d \right|^2 \nonumber \\ && \mbox{}
+ \frac{1}{2} g^2 \left( |h_u|^2 - |h_d|^2 - |l|^2 \right)^2
\end{eqnarray}
determines the dynamics of $h_u$, $h_d$ and $l$.
As $\phi$ rolls away from the origin, the $A$-term $A_\mu \lambda_\mu \phi^2 h_u h_d + \mbox{c.c.}$ and the cross term $\lambda_\nu^* {l^*}^2 h_u^* \lambda_\mu \phi^2 h_d + \mbox{c.c.}$ shift the minimum of $h_d$ to
\begin{equation} \label{MagHd}
h_d \simeq - \frac{\lambda_\mu^* {\phi^*}^2 \left( A_\mu^* h_u^* + \lambda_\nu l^2 h_u \right)}
{m_{H_d}^2 - \frac{1}{2} m_L^2 + \frac{1}{2} m_{H_u}^2}.
\end{equation}
The above equation is valid only for small $\phi$.
As $\phi$ and $h_d$ become larger, more terms become important and Eq.~(\ref{MagHd}) only gives a rough estimate.
This non-zero value of $h_d$ generated by $\phi$ is the key to generating a lepton asymmetry in our model, and is also crucial to maintain the stability of our ansatz as we will see in Section~\ref{stab}.
This was missed in Ref.~\cite{SKY}.

As $\phi$ approaches its minimum, the remaining terms become important.
The $A$-term $A_\phi \lambda_\phi \phi^4 + \mbox{c.c.}$ determines the phase of $\phi^4$, and the $A$-term $A_\mu \lambda_\mu \phi^2 h_u h_d + \mbox{c.c.}$ and the cross term $2 \lambda_\phi^* {\phi^*}^3 \lambda_\mu \phi h_u h_d + \mbox{c.c.}$ determine the phase of $\phi^2 h_u h_d$.
The cross term $\lambda_\mu^* {\phi^*}^2 h_d^* \lambda_\nu l^2 h_u + \mbox{c.c.}$ rotates the phase of $l h_u$, which was initially determined by the $A$-term $A_\nu \lambda_\nu l^2 h_u^2 + \mbox{c.c.}$, and at the same time $\left| \lambda_\mu \phi^2 h_u \right|^2$ gives $l h_u$ a positive mass squared bringing it back in towards the origin.\footnote{In the case that the flaton sector contains the axion, terms equivalent to the $A$-term $A_\phi \lambda_\phi \phi^4 + \mbox{c.c.}$ and the cross term $2 \lambda_\phi^* {\phi^*}^3 \lambda_\mu \phi h_u h_d + \mbox{c.c.}$ may be absent.
However, we see that the mechanism still works.}
The angular momentum of $l h_u$ generated in this way is our lepton number asymmetry.
Note that the cross term $\lambda_{\mu}^* {\phi}^{*2} h_d^* \lambda_{\nu}{l}^2 h_u + \mbox{c.c.}$ must be larger than, or at least comparable to, $A_{\nu} \lambda_{\nu} l^2 h_u^2 + \mbox{c.c.}$ to give an effective angular kick to the phase of $l h_u$.

After this, the dynamics becomes somewhat complicated.
We will analyze it numerically in Section~\ref{Sim} and discuss some of the relevant issues in Section~\ref{after}.

\section{Constraints and Stability} \label{CnS}

\subsection{Summary of constraints on the parameters} \label{Cons}

We assumed
\begin{equation}
m_\phi^2 < 0
\end{equation}
as is required for thermal inflation, and
\begin{equation}\label{LHumcon}
m_L^2 + m_{H_u}^2 < 0
\end{equation}
so that $L H_u$ becomes unstable near the end of the thermal inflation,
allowing the possibility for leptogenesis.
We will discuss the consistency and implications of Eq.~(\ref{LHumcon}) in Section~\ref{stab}.

We require any combination of $L H_u$ and $H_u H_d$ to be stable at the origin.
This gives the following constraints \cite{Munoz}.
\begin{equation} \label{HuHdstb}
m_{H_u}^2 + m_{H_d}^2 + 2 |\mu|^2 > 2 |B \mu|
\end{equation}
if $m_{L}^2 \geq m_{H_d}^2 + |\mu|^2 - |B \mu|$ so that pure $H_uH_d$ is the relevant direction, and
\begin{equation} \label{UFB2}
\left( m_{H_d}^2 + |\mu|^2 - m_{L}^2 \right)
\left( m_{L}^2 + m_{H_u}^2 + |\mu|^2 \right) > |B \mu|^2
\end{equation}
if $m_{L}^2 \leq m_{H_d}^2 + |\mu|^2 - |B \mu|$ so that some combination of $LH_u$ and $H_uH_d$ is the relevant direction.
If $B=0$, Eq.~(\ref{UFB2}) reduces to the stability constraint along the pure $LH_u$ direction
\begin{equation}
m_L^2 + m_{H_u}^2 + |\mu|^2 > 0.
\end{equation}
In our model, $\mu$ and $B$ are given by Eqs.~(\ref{mu}) and~(\ref{B}).
Note that Eqs.~(\ref{HuHdstb}) and~(\ref{UFB2}) are the well known MSSM constraints to avoid instability at large field values along a direction with nonzero $L$, $H_u$ or $H_d$ \cite{Munoz}.
Our inclusion of the neutrino mass term does not make things more unstable.

In addition to requiring that the origin is stable, we also have to worry whether the fields may get trapped in a minimum generated by the $A$-term $A_\nu \lambda_\nu l^2 h_u^2 + \mbox{c.c.}$.
It is not easy to determine explicit constraints that ensure that such a minimum does not exist for a general combination of $LH_u$ and $H_uH_d$ directions, but the constraint for the pure $LH_u$ direction is
\begin{equation}
m_{L}^2 + m_{H_u}^2 + |\mu|^2 > |A_{\nu}|^2 / 6.
\end{equation}
For the general case, we can check whether the A-term minima exist or not for a given set of parameters.
On the other hand, if the fields evade the minima dynamically, the constraints are not needed.
But it is not clear in our model.

Finally, we require
\begin{equation}
\left( m_{H_u}^2 + |\mu|^2 \right)
\left( m_{H_d}^2 + |\mu|^2 \right) < |B \mu|^2
\end{equation}
for correct electroweak symmetry breaking in our vacuum.

\subsection{Stability of our ansatz} \label{stab}

So far, we have analyzed our model and its dynamical behavior within the context of our ansatz that the only relevant fields are $H_u$, $H_d$, $L$ and $\phi$, assuming all other fields are zero.
However, our key constraint, Eq.~(\ref{LHumcon}),
\begin{equation}
m_L^2 + m_{H_u}^2 < 0
\end{equation}
violates the well known stability constraint for large field values along the directions of nonzero $\left( H_u, L, Q, d \right)$ or $\left( H_u, L, e \right)$ \cite{Komatsu,Munoz}.
This is dangerous because it means that very deep non-MSSM minima exist in our model.
Does this mean our assumption is invalid?
The answer will be no if we live in our vacuum and its life time is longer than the age of our universe.
Then two questions arise.
The first is whether our field configuration will dynamically settle down into our vacuum, as opposed to the deeper non-MSSM one.
If the answer to this question is yes, so that we live in a false vacuum, the second question is what is the life time of our vacuum.
To answer the second question first, the MSSM has a large enough region of parameter space in which our vacuum is cosmologically stable so that we are safe \cite{KLSe}.

Returning to the first question, one of the dangerous directions involves $Q_i^{1 \alpha}$ and $d_i^\alpha$.
For simplicity, we restrict to a single generation.
We can parameterize this single generation as
\begin{equation}
Q = \frac{1}{\sqrt{2}\,}
\left(\begin{array}{c} q \\ 0 \\ 0 \\\\ 0 \\ 0 \\ 0 \end{array}\right)
\comma
d = \frac{1}{\sqrt{2}\,}
\left(\begin{array}{c} q \\ 0 \\ 0 \end{array}\right)
\end{equation}
The superpotential becomes
\begin{equation}
W = \frac{1}{2} \lambda_d q^2 h_d + \lambda_\mu \phi^2 h_u h_d + \frac{1}{2} \lambda_\nu l^2 h_u^2 + \frac{1}{4} \lambda_\phi \phi^4
\end{equation}
and the potential
\begin{eqnarray}\label{Vwithq}
V & = & V_0 + m_{H_u}^2 |h_u|^2 + m_{H_d}^2 |h_d|^2 + \frac{1}{2} \left( m^2_Q + m^2_d \right) |q|^2 + m_L^2 |l|^2 + m_\phi^2 |\phi|^2 \nonumber \\ && \mbox{}
+ \left[ \frac{1}{2} A_q \lambda_d q^2 h_d + A_\mu \lambda_\mu \phi^2 h_u h_d + \frac{1}{2} A_\nu \lambda_\nu l^2 h_u^2 + \frac{1}{4} A_\phi \lambda_\phi \phi^4 + \mbox{c.c.} \right] \nonumber \\ && \mbox{}
+ \left| \lambda_\mu \phi^2 h_d + \lambda_\nu l^2 h_u \right|^2
+ \left| \frac{1}{2} \lambda_d q^2 + \lambda_\mu \phi^2 h_u \right|^2
+ \left| \lambda_d h_d q \right|^2 + \left| \lambda_\nu l h_u^2 \right|^2 \nonumber \\ && \mbox{}
+ \left| 2 \lambda_\mu \phi h_u h_d + \lambda_\phi \phi^3 \right|^2
+ \frac{1}{2} g^2 \left( |h_u|^2 + \frac{1}{2} |q|^2 - |h_d|^2 - |l|^2 \right)^2.
\end{eqnarray}
As $\phi$ rolls away from the origin at the end of thermal inflation, there is a danger that the term $\left| \lambda_d q^2 / 2 + \lambda_\mu \phi^2 h_u \right|^2$ may destabilize $q$ rather than bring $h_u$ back in to the origin.
However, as discussed in Section~\ref{end}, the A-term $A_\mu \lambda_\mu \phi^2 h_u h_d + \mbox{c.c.}$ and cross term $\lambda_\nu^* {l^*}^2 h_u^* \lambda_\mu \phi^2 h_d + \mbox{c.c.}$ induce a non-zero value for $h_d$, which tends to stabilize $q$ via the term  $\left| \lambda_d h_d q \right|^2$.
Thus, ignoring finite temperature effects which will also help to hold $q$ at zero, $q$ will be stable if
\begin{equation} \label{Qdm}
\frac{1}{2} \left( m^2_Q + m^2_d \right)
- \left| \lambda_d h_d \right| \left| A_q + \frac{\lambda_\mu^* {\phi^*}^2 h_u^* h_d^*}{|h_d|^2} \right|
+ \left| \lambda_d h_d \right|^2
+ \frac{1}{2} g^2 \left( |h_u|^2 - |h_d|^2 - |l|^2 \right) > 0
\end{equation}
The D-term contribution is not expected to be large because the field configuration will follow a direction close to D-flat, and in any case would be expected to be positive, and the second term is not expected to dominate as Eq.~(\ref{MagHd}) makes it roughly the geometric mean between the first and third terms.
Thus one expects that Eq.~(\ref{Qdm}) will be satisfied, i.e. $q$ is stable.
We confirm this estimate using numerical simulations in the next Section.

For the dangerous lepton direction
\begin{equation}
L = \left(\begin{array}{c} \frac{1}{\sqrt{2}\,} \nu \\ l \end{array}\right)
\comma
e = \frac{1}{\sqrt{2}\,} \nu
\end{equation}
we can apply the same argument as above with the substitutions
\begin{eqnarray}
q \rightarrow \nu, & \lambda_d \rightarrow \lambda_e, & (m_Q^2,
m_d^2) \rightarrow (m_L^2, m_e^2).
\end{eqnarray}

\section{Numerical Simulation of the Dynamics} \label{Sim}

In this section we use numerical simulations of the homogeneous field dynamics to see the generation of the lepton asymmetry and check the stability of our ansatz.

For our initial conditions, we take $\phi = h_d = 0$, and $l$ and $h_u$ at the minimum of their potential.
We take $q$ slightly displaced from the origin to test its stability.
We then displace $\phi$ slightly from the origin at various angles, and follow the dynamics.
To crudely mimic the transfer of energy from the initial homogeneous mode of $\phi$ to gradient energy of $\phi$, which causes the homogeneous mode of $\phi$ to settle towards its minima, we add artificial damping acting on $\phi$.
Note that Hubble damping is negligible.

There are various phases in our potential, but only certain combinations are physical.
To make things more explicit, by field rotations we can choose
\begin{eqnarray}
\arg \left( A_X \lambda_X \right) = 0.
\end{eqnarray}
The remaining physical phases are
$\arg \left( - \lambda_\mu^* \lambda_\nu \right)$ and
$\arg \left( - \lambda_\mu^* \lambda_\phi \right)$,
with $CP$ conserving values being $0$ and $\pi$.

We define the rescaled variables
\begin{eqnarray}
&&
\hat{t} = m t \comma
\hat{m}_X^2 = \frac{m_X^2}{m^2} = \mathcal{O}(1) \comma
\hat{A}_X = \frac{A_X}{m} = \mathcal{O}(1) \comma
\\ &&
\hat{h}_u = \frac{h_u}{M_\mathrm{AD}} \comma
\hat{h}_d = \frac{h_d}{M_\mathrm{AD}} \comma
\hat{q} = \frac{q}{M_\mathrm{AD}} \comma
\hat{l} = \frac{l}{M_\mathrm{AD}} \comma
\hat{\phi} = \frac{\phi}{M_\mathrm{TI}} \comma
\\ &&
\hat{\lambda}_\mu = \frac{\lambda_\mu M_\mathrm{TI}^2}{m} = \mathcal{O}(1) \comma
\hat{\lambda}_\nu = \frac{\lambda_\nu M_\mathrm{AD}^2}{m} = \mathcal{O}(1) \comma
\hat{\lambda}_\phi = \frac{\lambda_\phi M_\mathrm{TI}^2}{m} = \mathcal{O}(1) \comma
\\ &&
\hat{\lambda}_H = \frac{\lambda_H M_\mathrm{AD}^2}{m} = \mathcal{O}(10^{-10}) \ \mbox{to} \ \mathcal{O}(10^{-2}) \comma
\\ &&
\hat{g} = \frac{g M_\mathrm{AD}}{m} = \mathcal{O}(10^5) \ \mbox{to}\ \mathcal{O}(10^{7}) \comma
\hat{\lambda}_d = \frac{\lambda_d M_\mathrm{AD}}{m} = \mathcal{O}(1) \ \mbox{to}\ \mathcal{O}(10^{7})
\end{eqnarray}
where
\begin{equation}
\frac{M_\mathrm{AD}^2}{m} = \mathcal{O}(10^{-6}) \ \mbox{to}\
\mathcal{O}(10^{-2}) \comma \frac{M_\mathrm{TI}^2}{m} =
\mathcal{O}(1) \ \mbox{to}\ \mathcal{O}(10^4).
\end{equation}
The potential can be divided into two parts
\begin{equation}
V = V_\mathrm{TI}(\phi) + V_\mathrm{AD}(h_u,h_d,q,l,\phi)
\end{equation}
as in Eqs.~(\ref{VTI}) and~(\ref{VAD}), and rescaled as
\begin{equation}
\hat{V}_\mathrm{TI} = \frac{V_\mathrm{TI}}{m^2 M_\mathrm{TI}^2} \comma
\hat{V}_\mathrm{AD} = \frac{V_\mathrm{AD}}{m^2 M_\mathrm{AD}^2}.
\end{equation}
The rescaled Lagrangian is
\begin{eqnarray}\label{energy}
\hat{\mathcal{L}} & = & \frac{\mathcal{L}}{m^2 M_\mathrm{TI}^2} \\
& = & \hat{K}_\mathrm{TI} \left( \hat{\partial}\hat{\phi} \right)
+ \hat{V}_\mathrm{TI} \left( \hat{\phi} \right)
+ \left( \frac{M_\mathrm{AD}}{M_\mathrm{TI}} \right)^2
\left[ \hat{K}_\mathrm{AD} \left( \hat{\partial}\hat{h}_u, \hat{\partial}\hat{h}_d, \hat{\partial}\hat{q}, \hat{\partial}\hat{l} \right)
+ \hat{V}_\mathrm{AD} \left( \hat{h}_u, \hat{h}_d, \hat{q}, \hat{l}, \hat{\phi} \right) \right]
\nonumber \\
\end{eqnarray}
where the $K$'s are the kinetic terms.
Note that $V_\mathrm{AD}$ only has a small effect on the dynamics
of $\phi$ because
\begin{equation} \label{MadMti}
\left( \frac{M_\mathrm{AD}}{M_\mathrm{TI}} \right)^2
= \mathcal{O}(10^{-10}) \mbox{ to } \mathcal{O}(10^{-2}).
\end{equation}

The equation of motion for $\phi$ is given by
\begin{equation}
\frac{d^2\hat{\phi}}{d\hat{t}^2} + \Gamma \frac{d\hat{\phi}}{d\hat{t}}
+ \frac{\hat{\partial}\hat{V}_\mathrm{TI}}{\hat{\partial}\hat{\phi}^*} = 0
\end{equation}
with $\Gamma$ our artificial damping.
We do not include any damping for the AD fields as it is less clear what form it takes.

Our numerical simulations show that $q$ is stable.
Thus the dynamics ensures that we do not fall into a deeper non-MSSM minimum, and our ansatz and key assumption, Eq.~(\ref{LHumcon}), are consistent.

\begin{figure}[p]
\centerline{\epsfxsize=4 in \epsfbox{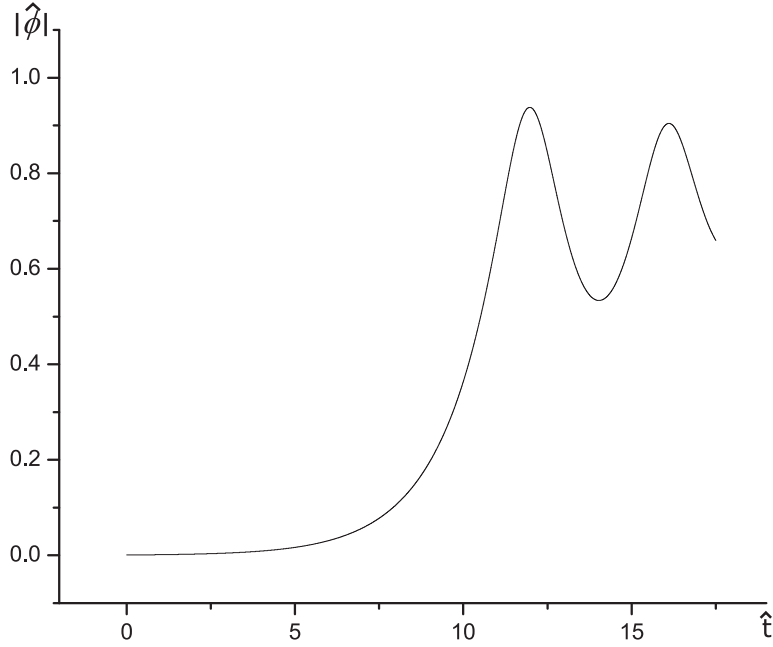}}
\caption{\label{phi} $|\hat\phi|$ as a function of time, for
$\hat{m}_{\phi}^2 = -0.5$, $|\hat{A}_\phi| = 1.0$,
$|\hat{\lambda}_\phi| = 1.0$, initial phase $\arg\phi_\mathrm{i} =
3 \pi/20$, and artificial damping $\Gamma = 0.2$.}
\end{figure}

\begin{figure}[p]
\centerline{\epsfxsize=4 in \epsfbox{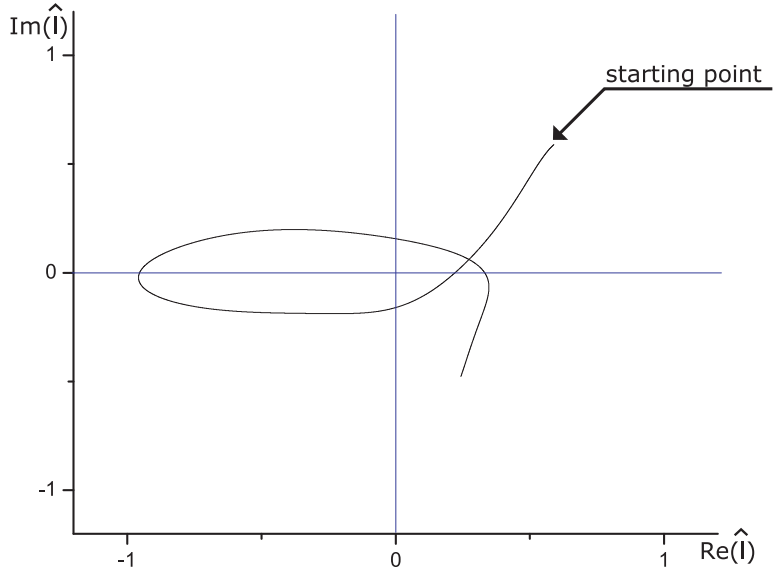}}
\caption{\label{Ldyn} Dynamics of $\hat{l}$, for the same parameters as Figure~\ref{phi}, and
$\hat{m}_{H_u}^2 = -2.5$,
$\hat{m}_{H_d}^2 = 2.4$,
$\hat{m}_{L}^2 = 1.0$,
$|\hat{A}_\nu| = 1.0$,
$|\hat{A}_\mu| = 1.0$,
$|\hat{\lambda}_\nu| = 1.0$,
$|\hat{\lambda}_\mu| = 2.5$,
$\hat{g} = 10^3$, $\arg \left( - \lambda_\mu^* \lambda_\nu \right) = \pi + \pi / 120$, and
$\arg \left( - \lambda_\mu^* \lambda_\phi \right) = \pi + \pi / 300$.
The correct value of $g$ is $\hat{g} \sim 10^{6}$, but the value we used is large enough to make little difference and is easier to handle numerically.
Angular momentum is generated, but, as we did not include damping for the AD fields, we do not expect it to survive in our simulation.}
\end{figure}

\begin{figure}[ht]
\centerline{\epsfxsize=4 in \epsfbox{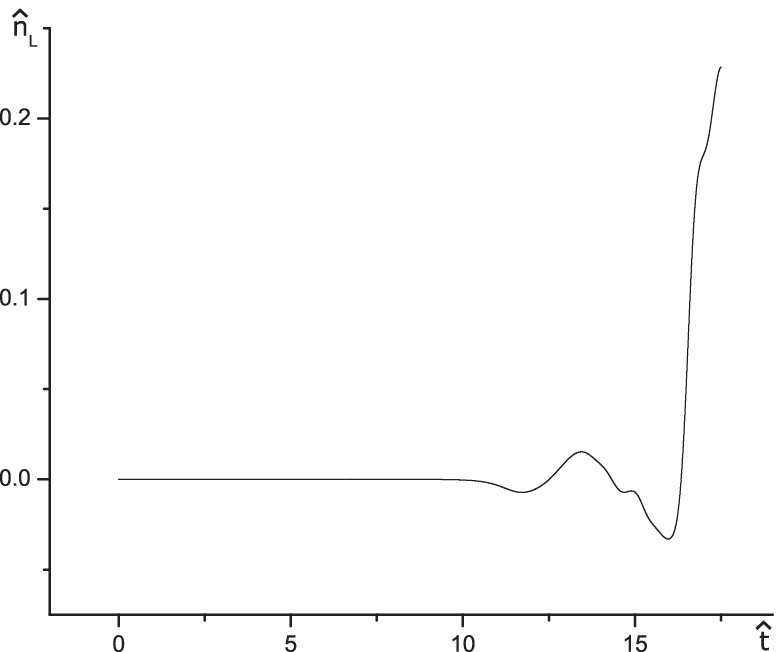}}
\caption{\label{av} The lepton number asymmetry,
$\hat{n}_L = \mathrm{Im} \left( \hat{l}^* \, \hat{\dot{l}} \right)$, averaged over the initial phase of $\phi$, as a function of time, for the same parameters as Figures~\ref{phi} and~\ref{Ldyn}.
A lepton asymmetry is generated, but, as we did not include damping for the AD fields, we do not expect it to survive in our simulation.}
\end{figure}

Some of our numerical results are shown in Figures~\ref{phi}, \ref{Ldyn} and~\ref{av}.

\section{Preserving the Lepton Asymmetry} \label{after}

In the previous sections we showed that a lepton asymmetry can be generated in our model at the end of thermal inflation.
However, there is a danger that this lepton number will be washed out by the continued action of the lepton violating operators.
To avoid this, we need the amplitude of the lepton field to decrease, taking it into the conservative region of the potential.
Therefore, some form of damping for the homogeneous mode of the lepton field is required to preserve the asymmetry.
Note that Hubble damping, which is usually used in AD baryogenesis scenarios, is negligible.

There are several sources of damping.
The first we can expect is the transfer of energy from the homogeneous mode to inhomogeneous modes, i.e. the build up of gradient energy, which is often called preheating.
It is clear that a substantial amount of energy would be transferred in this way in the first few oscillations, but it is not clear how long it would take this process to reduce the homogeneous field amplitude by a substantial factor.
Friction induced by the thermal bath and decay are other sources of damping.
We don't attempt to analyze these damping processes in this paper.

Although we do not understand the details of the decay of the AD fields, we do know that it will complete while the flaton energy is still dominant, and will release enough thermal energy to restore the electroweak symmetry.
Assuming our lepton asymmetry survives until this stage, it will be converted to a baryon asymmetry by the usual electroweak sphaleron processes.
Finally, as the temperature drops to the GeV scale, the flaton decay completes, releasing substantial entropy which dilutes the baryon asymmetry.

Thus, the final baryon asymmetry is given by
\begin{equation}
\frac{n_B}{s} \sim \frac{n_B}{n_\phi} \frac{T_\phi}{m_\phi}
\sim \frac{n_L}{n_\mathrm{AD}} \frac{n_\mathrm{AD}}{n_\phi} \frac{T_\phi}{m_\phi}
\sim \frac{n_L}{n_\mathrm{AD}} \left(\frac{M_\mathrm{AD}}{M_\mathrm{TI}}\right)^2 \frac{T_\phi}{m_\phi}.
\end{equation}
As in Eq.~(\ref{MadMti}), we expect
\begin{equation}
\left(\frac{M_\mathrm{AD}}{M_\mathrm{TI}}\right)^2 \sim 10^{-10} \mbox{ to } 10^{-2},
\end{equation}
and as $m_\phi \sim m_\mathrm{EW}$ and $T_\phi \sim 10^{-1}$ to $10$ GeV, we expect
\begin{equation}
\frac{T_\phi}{m_\phi} \sim 10^{-3} \mbox{ to } 10^{-1}.
\end{equation}
Therefore, to get
\begin{equation}
\frac{n_B}{s} \sim 10^{-10}
\end{equation}
we require
\begin{equation}
\frac{n_L}{n_\mathrm{AD}} \gtrsim 10^{-7}
\end{equation}
with perhaps
\begin{equation}
\frac{n_L}{n_\mathrm{AD}} \sim 10^{-2}
\end{equation}
being a central value.

\section{Conclusion} \label{dis}

Our baryogenesis scenario emerges from a fairly minimal extension
of the MSSM with superpotential of the form
\begin{equation}
W = \lambda_u Q H_u u + \lambda_d Q H_d d + \lambda_e L H_d e
+ \lambda_\mu \phi^2 H_u H_d + \lambda_\nu (L H_u)^2
+ \lambda_\phi \phi^4 + \lambda_\chi \phi \bar\chi \chi.
\end{equation}
$\phi$ is a flaton field whose potential drives thermal inflation, solving the moduli problem.
$\phi$ also naturally generates the $\mu$-term of the MSSM when it acquires its intermediate scale vacuum expectation value.
$\chi$ and $\bar\chi$ represent some additional SU(5) multiplets needed to hold the flaton at the origin during the thermal inflation.
After the thermal inflation they acquire intermediate scale masses from the flaton vacuum expectation value.
Their effect on the renormalization group running of MSSM parameters from the GUT scale might provide a signature of our model.
One could also consider embedding the Peccei-Quinn axion in the flaton sector.

Our model makes the key assumption that
\begin{equation}
m_{L}^2 + m_{H_u}^2 < 0
\end{equation}
for our AD fields to become unstable near the end of thermal inflation.
This violates a well known stability constraint \cite{Munoz,Komatsu}, implying that there exists a deep non-MSSM minimum in our model.
This is potentially dangerous.
However, we showed explicitly that the dynamics does not cause the fields to become trapped in this deeper minimum.
It is also known that the time scale for quantum tunnelling from our vacuum to this deeper minimum is larger than the age of our universe for a wide range of parameters \cite{KLSe}.
So the above assumption is phenomenologically consistent and becomes a clear signature of our model which can be checked in future particle accelerator experiments.

The outline of our baryogenesis scenario is as follows.
We start with all fields at the origin during thermal inflation.
As the temperature drops, near the end of thermal inflation, the $LH_u$ flat direction rolls away from the origin.
Then, $\phi$ rolls away from the origin, ending the thermal inflation and inducing a nonzero value for $H_d$.
This non-zero value for $H_d$ stabilizes some potentially dangerous quark and lepton field directions, shielding the dynamics from the deep non-MSSM minima discussed above.
When $\phi$ reaches of order its vacuum expectation value, the back-reaction of $H_d$ induces a lepton number violating cross-term which rotates the phase of $LH_u$ generating a lepton number asymmetry.
Simultaneously, $L$, $H_u$ and $H_d$ are brought back in towards the origin due to the $\mu$-term generated by $\phi$.
We assume the generated asymmetry is conserved due to damping of the amplitude of the AD field oscillations by preheating, thermal friction and decay processes.
While the oscillating flaton still dominates the energy density of the universe, the decay of the AD fields completes, partially reheating the universe to a temperature high enough to restore the electroweak symmetry.
Sphaleron processes then convert the lepton asymmetry into a baryon asymmetry.
Finally, the flaton decay completes at a temperature of order GeV, diluting the baryon asymmetry to the observed value.

As described above, despite its minimal and strongly constrained structure, our model has good features for baryogenesis providing rich physical content.
This is a very attractive feature of our model.

One key topic for future work is a proper analysis of the damping and decay of the AD fields, which is crucial to the conservation of the generated lepton asymmetry.
Our numerical simulation was only intended to model the generation of our asymmetry and check the stability of our ansatz, and so only simulated homogeneous fields.
Thus it could not address the damping and decay issues.
Therefore, to complete the story, a full analysis of the damping and decay processes and an inhomogeneous simulation including decay products are required.
We postpone this challenging task to future work.

\subsection*{Acknowledgements}
EDS thanks Masahide Yamaguchi for collaboration on an earlier related project,
John Terning for a stimulating talk, and the SF02 Cosmology Summer Workshop for hospitality.
EDS and WIP thank Patrick Greene for collaboration at an earlier stage of this project.
KK thanks Joanne D. Cohn for her continuous encouragement.
DJ, EDS and WIP were supported in part by the Astrophysical Research Center for the Structure and Evolution of the Cosmos funded by the Korea Science and Engineering Foundation and the Korean Ministry of Science, the Korea Research Foundation grant KRF PBRG 2002-070-C00022, and Brain Korea 21.
The work of KK was partially supported by NSF grant AST-0205935.

\end{document}